# The Deep Physics Behind the Second Law: Information and Energy As Independent Forms of Bookkeeping

**Todd L. Duncan**

Center for Science Education, Portland State University, P.O. Box 751, Portland, OR 97207, USA. Tel. (503) 805-7860; E-mail duncant@pdx.edu.

**Jack S. Semura**

Department of Physics, Portland State University, P.O. Box 751, Portland, OR 97207, USA. Tel. (503) 725-4229; E-mail semuraj@pdx.edu.

**Abstract:** Even after over 150 years of discussion, the interpretation of the second law of thermodynamics continues to be a source of confusion and controversy in physics. This confusion has been accentuated by recent challenges to the second law and by the difficulty in many cases of clarifying which formulation is threatened and how serious the implications of a successful challenge would be. To help bring clarity and consistency to the analysis of these challenges, the aim of this paper is to suggest a simple formulation of the deep physics of the second law, and to point out how such a statement might help us organize the challenges by level of seriousness. We pursue the notion that the second law is ultimately a restriction operating directly on the dynamics of information, so the existence of this law can be traced to the need for a system of "information bookkeeping" that is independent of the bookkeeping for energy. Energy and information are related but independent, so the dynamical restrictions for one cannot be derived from those for the other. From this perspective, we also suggest the possibility that the foundation of the second law may be linked to the finite capacity of nature to store information about its own state.

## Introduction

The second law of thermodynamics has a long and contentious history in physics. A recent string of challenges to the absolute status of the law has brought greater urgency to the problem of clarifying the basis for our strong belief in the law and our understanding of what it means to violate it [1]. We cannot clearly identify whether the essential second law is threatened at a fundamental level unless we know which version of the second law we are challenging and how fundamental that version is.

This raises the question: Is it possible to put forth a strong statement of the "deep physics" (borrowing the phrase used by Scully at QLSL 2002 [2]) of the second law, which describes the underlying principle of nature for which the individual formulations of the law are special cases? We would like to be able to clearly articulate what essential feature of nature one would have to change in order to 'turn off' the second law. It seems likely that traditional statements of the second law, formulated with reference to particular circumstances, play the role of specific representations of a deeper more fundamental law. So it is possible that some "violations" of particular statements of the second law may only be indications that one particular statement does not quite accurately express the underlying principle of nature, and needs to be refined. On the other hand, some challenges may indeed pose a threat to a deeper version of the second law. With a clear statement of this underlying principle in hand, we would be in a better position to classify challenges to it in terms of seriousness. If they challenge the deep physics, then we are concerned, but if they challenge only a particular more limited statement, then we would look to see why this situation does not fit that special statement, and what essential feature such a statement fails to capture about the deeper second law.

Our aim here is to help set a context for discussion, by articulating one way to think about the underlying meaning of the second law and pointing out some directions that may be fruitful to pursue. We begin by suggesting a way to state the underlying property of nature that is responsible for the second law, formulated in terms of information which provides the most natural expression. We also suggest that the dynamics of information are fundamentally independent of the dynamics of energy (*i.e.*, that there are two kinds of independent "bookkeeping" that govern information and energy). We then show how this information statement connects to and underlies traditional thermodynamic statements, and finally suggest a classification of second law challenges within this context.

For convenience, we summarize here the core points to be made in this paper:

1) The second law is a law about information – it operates at the level of information, not energy, and hence requires a separate bookkeeping system for information in order to impose the law. An implication of this line of thinking is that a description of physics based only on energy bookkeeping (*i.e.* a Hamiltonian/Lagrangian mechanics with energy as the generator of time evolution for the system) is incomplete. Of course the energy and information bookkeeping systems must be *consistent* with each other, but the dynamics of information is independent and equally necessary to describe the world.

2) In terms of information, the second law says that information is truly and fundamentally erased by some processes, so that once erased, that information cannot be recovered.

3) We show that our statement contains the traditional statements of the second law; i.e. there is a clear link between the direction of heat flow and loss of information.

4) This perspective leads to a useful classification of potential violations: those claimed second law challenges which amount to actually recovering erased information, and those in which the information was never really lost in the first place.

**A proposed statement of the deep physics, in terms of information**

Since our basic argument is that the second law is most simply and fundamentally a law about information, we begin by suggesting a formulation in terms of information. (Here and throughout this paper, information means *classical* information.) In a later section we will connect this formulation to the traditional statements of the second law. We suggest that the second law as we observe it stems from the fact that

*(classical) information about the state of a system is fundamentally erased by many processes, and once truly lost, that information cannot be reliably recovered.*

"Reliable recovery" here means information recovery with a success rate higher than that governed by a statistical treatment – *i.e.* higher than one could achieve by "guessing."

Stated in this form, the reason we experience a second law of thermodynamics is essentially the same as the reason we cannot recover the high-resolution details of a digital image after we have saved the image as a lower-resolution file. If we start with a detailed, high-resolution image, we can save memory by throwing away information and storing it as a lower-resolution image. But once we have done this, we cannot ever recover the detail information that was contained in the original image (assuming that there is no redundancy in the image file [3] and also assuming that the file itself is the only information we keep – we don't keep track also of the previous, high-resolution file, because that would require additional memory). For example, if the resolution is reduced by a factor of 4, then 4 pixels from the original image have been compressed into one to make the smaller file. So the information about what was in each of those original pixels *no longer exists*, and cannot be recovered from the smaller file.

**Connection to traditional thermodynamic formulations of the second law**

The standard thermodynamic statements can be connected to the information formulation by noticing that traditional statements express the limited accessibility of energy once it is in the form of heat. So from a particular observer's point of view, the second law essentially states that: 1) Energy tends to flow into unobservable degrees of freedom (those which we are incapable of manipulating directly); 2) Once there, the energy cannot be recovered into observable degrees of freedom without accompanying changes that cause even more energy to be transferred from observable to unobservable degrees of freedom.

The limited accessibility of energy in the form of heat can be treated as a loss of information about the detailed state of the system. We cannot transfer all of that energy into some other degree of freedom because we do not know exactly where it is and don't have direct control over it.

This connection between the information formulation of the second law and traditional statements stems from the following core ideas:

*Information is physical* – Central to our argument is the fundamental link between information and entropy. This important idea emerged from Szilard's 1929 [4] analysis of the Maxwell Demon problem, and has been formulated more sharply as Laundauer's principle [5] which directly and formally links entropy to information. By pointing out that information cannot exist as a pure abstraction, but must be embodied in a physical state and thus subject to the physical laws controlling the state of the system in which the information is stored, Landauer argued that the erasure of information must generate thermodynamic entropy. Specifically, erasing one bit of information must increase entropy by an amount kln2, or with erasing written as a negative change in information:

$$\Delta S = -k \ln 2 \Delta I. \qquad (1)$$

*Temperature connects information and energy* – Specifically, kTln2 is the proportionality between energy and information, obtained by combining (1) with the familiar thermodynamic relation, $\Delta Q = T \Delta S$. Thus

$$\Delta Q = -kT \ln 2 \Delta I. \qquad (2)$$

This is just another way of stating Landauer's principle: erasing one bit of information dissipates at least kTln2 of energy as heat.

So when energy is exchanged between two systems, information is also exchanged (see Figure 1), but the dynamics of energy exchange does not uniquely determine the information exchanged. For the same amount of energy, different amounts of information can flow in or out of a system. This is why temperature is necessary as a parameter to describe the system – it specifies the relationship between information and energy, as given in Equation 2.

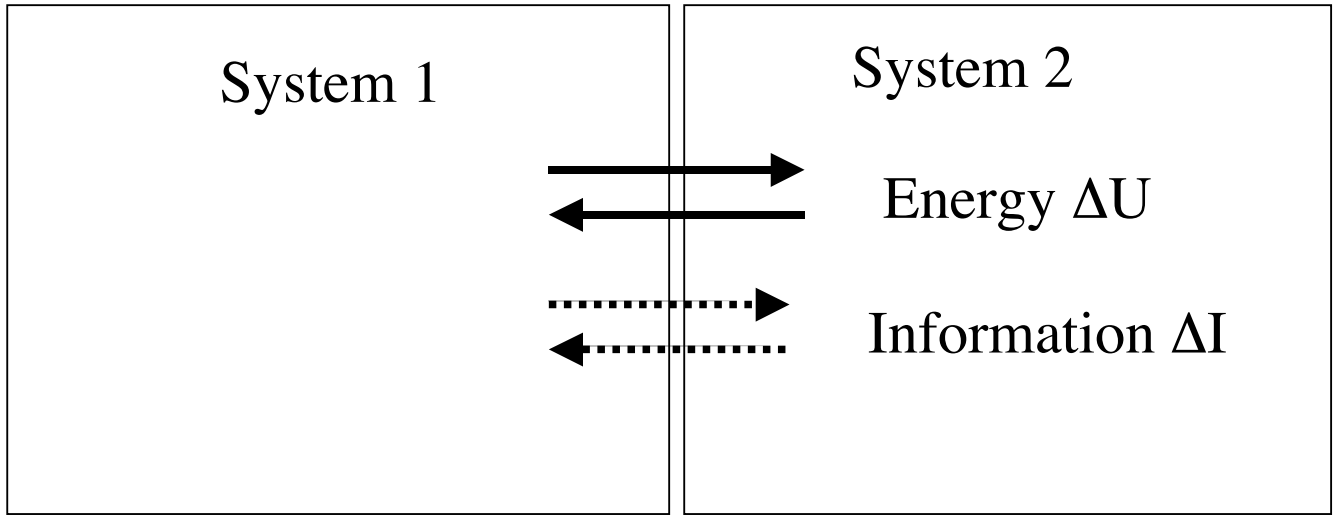

**Figure 1.** When energy and information are exchanged between two systems, the dynamics of energy exchange does not uniquely determine the information exchanged.

With this background, we can make the direct link between heat flow and information loss. Consider an exchange of heat between a hot system at temperature $T_H$ and a cooler system at temperature $T_C$. If the net heat flow $\Delta Q$ is from hot to cold as required by the Clausius form of the second law, this is equivalent to a net erasure of information from the world. To see this requires only a restatement of the standard textbook treatment describing the change in accessible states in the two systems. When $\Delta Q$ flows out of the system at $T_H$, the corresponding change in information is

$$\Delta I_H = \frac{\Delta Q}{kT_H \ln 2}. \qquad (3)$$

That is, the state of the hot system is more well-defined because it has less energy and hence fewer accessible states, so information has been *gained* about the state of that system. At the same time, we lose information about the state of the cooler system:

$$\Delta I_C = \frac{-\Delta Q}{kT_C \ln 2}. \qquad (4)$$

Since $T_H > T_C$, $\Delta I_{total} = \Delta I_H + \Delta I_C < 0$, and there is a net loss of information from the total system. Thus if we ask the question, "Why does heat flow from hot to cold," the answer in this context is that if heat flowed from cold to hot, information that had been erased would have to be recovered. So heat flowing from hot to cold would be analogous to recovering a high resolution image using only the information stored in the low resolution version.

We should point out that our approach differs from the maximum entropy formalism as developed by Jaynes and others [6]. In maximum entropy the central quantity is the information *entropy* S' defined in terms of the *average* information S' = <I>, where the average is taken over an inferred probability ensemble. In the present paper, however, we have used Landauer's principle to recast the second law directly in terms of *information* rather than in terms of *average information*.

**Information dynamics as fundamental and independent of energy dynamics**

The second law seems very natural when expressed in terms of information. Its underlying basis in this form lies in the simple fact that information which has been truly erased cannot be recovered, any more than we can recover details of an image when the information about those details is no longer stored in any form. But in order to state the law in this simple and natural form, note that we are treating *information as a fundamental quantity* so that the restriction imposed by the second law is described directly in terms of information. Information can be erased, and in that process something real and fundamental has changed about the state of a system. Energy dynamics alone cannot describe this change; following the flow of energy does not tell you directly whether there are distinguishable states in which information is stored. Thus while information is related to energy this version of the second law implies that energy and information bookkeeping are truly separate. An implication of this line of thinking is that a description of physics based only on energy bookkeeping (*i.e.* a Hamiltonian/Lagrangian mechanics with energy as the generator of time evolution for the system) is incomplete. The dynamics of energy exchange does not always pin down *which* states are gaining or giving up the energy. Of course the energy and information bookkeeping must be *consistent* with each other, but the dynamics of information is independent and equally necessary to describe the world, and the information description cannot be reduced to the energy description.

There is much to be explored along this line of thinking. This framework may help bring clarity to some of the issues involved in trying to "derive" the second law from microscopic physics, or trying to discuss the incompatibility of it with microphysics. This framework also invites the question, "*Why* is information erased as time passes?" One possibility worth suggesting is the notion that nature has a finite capacity to store information about its own state, so that information must be discarded as time evolves in order to make room to store the state of things in the next moment of time.

**Proposed classification of second law challenges**

Within the context of this information statement of the second law, we can categorize challenges to the law as follows:

**Class I**: Challenges that are true violations of the deep physics of the second law, because they actually recover lost classical information.

**Class II**: Apparent violations in which the information was not actually lost, but in fact was still stored in a form that our bookkeeping did not account for.

Note that well-posed challenges in both classes are interesting and offer the opportunity to deepen our understanding of nature. Class I challenges, if upheld, would obviously lead to significant revisions in our understanding of physics. But from this perspective, they specifically offer the opportunity to probe the question of why information tends to be erased as time passes, and to gain insight into how new information is generated. Class II challenges are particularly interesting within this information context, because this framework makes it easier to see that they are clearly not a threat to the physics behind the second law. Thus it may be easier to investigate these challenges with an open mind. Several of the recent challenges seem to fall into Class II [6]. These situations might be interpreted as cases where quantum entanglement provides a channel in which information about the state of the system is stored, but since this channel is not usually accounted for in traditional statements of the second law, these formulations treat this information as if it has been lost. In other words, there are channels for information flow that have been overlooked, so it appears that the second law has been violated (and in fact that particular limited formulation of the law *has* been violated), but really we just didn't have entries in our books to keep track of where all the information was. From this perspective the second law opens an avenue for progress in understanding forms of information analogous to that provided by the first law for forms of energy. (An example is the discovery of the neutrino as an unaccounted for channel for storing information, when it appeared that energy conservation had been violated.) This provides a fruitful framework for thinking about these challenges – the deep physics related to information loss is intact, but we learn new ways to store information which must be included in our accounting.

Finally, attempting to categorize second law challenges as Class I or Class II might provide a useful structure in which to analyze them. For example, Sheehan et al. [7] have presented an intriguing set of challenges which defy standard attempts to explain away the apparent second law violation. A valuable first step in understanding and resolving such challenges might be to trace the flow of information and isolate the process which appears to be recovering erased information about the system.

*Acknowledgments*: The authors would like to dedicate this work to Patty Jeanne Semura.